\documentclass[dvips]{acta}
\usepackage{supertabular,lscape,epsfig}
\usepackage{longtable}
\usepackage{rotating}
\usepackage{amssymb}
\usepackage{amsmath}
\usepackage{wasysym}
\usepackage{color}

\SetPages{0}{0}

\SetVol{58}{2008}

\begin{document}

\begin{Titlepage}
\Title{Light-curves of symbiotic stars in massive photometric surveys I: D-type systems}
\Author{M.~~G~r~o~m~a~d~z~k~i$^1$,~~J.~~M~i~k~o~{\l}~a~j~e~w~s~k~a$^{1}$,\\
P.~~W~h~i~t~e~l~o~c~k$^{2,3}$,~~F.~~M~a~r~a~n~g$^{2}$}
{$^1$N. Copernicus Astronomical Center, Bartycka 18, PL-00-716 Warsaw, Poland\\
e-mail: {\small marg,mikolaj@camk.edu.pl}\\
$^2$South African Astronomical Observatory, 
P.O. Box 9, Observatory, 7935, South Africa\\
e-mail: {\small paw,fm@saao.ac.za}\\
$^3$National Astrophysics and Space Science Programme,
Department of Astronomy, University of Cape
Town, Rondebosch, 7701, South Africa}

\Received{Month Day, Year}
\end{Titlepage} 

\Abstract{ASAS, MACHO, OGLE and SAAO $JHKL$ light curves of 13 stars, that
have at some time been classified as D-type symbiotics, are analysed.  Most
of the near-IR light-curves that have been monitored over many years
show long-term changes due to variable dust obscuration, in addition to the
stellar pulsation. The distances to these objects are derived from the
period-luminosity relation and estimates of the mass-loss rates made from
the $K_{0}-[12]$ colour.

We reclassify AS~245 as an S-type symbiotic, with a semi-regular cool
component with a pulsation period of about one year. The periods of the
large amplitude pulsations of SS73\,38 (463 days), AS~210 (423 days) and
H2-38 (395 days) are estimated for the first time, confirming that they are
symbiotic Miras.

A comparison of the symbiotic Miras with normal Miras of similar
pulsation period shows that the symbiotic stars have on average higher
values of $K_{0}-[12]$. This may indicate that they have higher mass-loss
rates, or more likely that the dust which is being lost by the Mira is
trapped within the binary system.   

} {{binaries: symbiotic -- Stars: individual: $o$~Cet, RX~Pup,
V366~Car, BI~Cru, SS73~38, V347~Nor, AS~210, AS~245, H~2-38, RR~Tel, R~Aqr,
StHA~55, V335~Vul -- surveys}}

\Section{Introduction}

Symbiotic stars are long-period interacting binary systems, in which an
evolved red giant transfers material onto its much hotter companion, which
in most systems is a white dwarf. Based on their near-IR characteristics,
symbiotic stars divide into two main classes (Allen 1982) depending whether
the colours are stellar (S-type) or indicate a thick dust shell (D-type).
The majority ($\sim$80\%) of catalogued systems are S-type and have near-IR
colours consistent with cool stellar photosphere temperatures of $\sim
3500-4000$\,K. Most of them have orbital periods $\sim 500-1000$ days ({\it
e.g.} Miko{\l}ajewska 2008). The near-IR colours of D-type systems indicate
the presence of a dust shell which obscures the star and re-emits at longer
wavelengths. IR photometric monitoring has shown that these D-type systems
have large amplitude variations and that they contain Mira variables with
pulsation periods in the range 300--600 days; they are often called
symbiotic Miras (Whitelock 1987). Since they must accommodate the Mira with
its dust shell, these D-type systems should have much longer orbital periods
than the S-types, a few tens of years and more. The latest review of
symbiotic Miras and a comparison with normal Miras can be found in Whitelock
(2003).

Light curves of symbiotic stars reflect the very complex behaviour of these
systems.  They show high and low activity stages, flickering, nova-like
outbursts originating from the hot component (S \& D types), eclipses,
ellipsoidal variability connected with orbital motion (S-type), radial
pulsations (all D-type and some S-type) and semi-regular variation (S-type)
of the cool component, long-term dust obscuration (mostly D-type) and other
types of variability (Miko{\l}ajewska 2001).

In this paper we analyse the light curves of 13 objects that have at
some time been classified as D-type symbiotics. The light-curves were
provided by massive photometry surveys such as ASAS, OGLE, MACHO and near-IR
monitoring at SAAO.

\Section{Data}

Belczy{\'n}ski \etal (2000) listed coordinates for symbiotic stars, but many
of these are not sufficiently accurate to identify the symbiotics
unambiguously. So we first identified the 2MASS counterparts using the
existing finding charts and the Aladin Java graphics interface running at
the CDS in Strasbourg. This works well because symbiotic stars, which have
the near-IR colours of late-type giants, are intrinsically bright in $JHK$.
The 2MASS coordinates were then used to identify symbiotic stars in the
OGLE, MACHO and ASAS databases.

For $o$~Cet, RX~Pup, V366~Car, BI~Cru, SS73~38, AS~210, RR~Tel, R~Aqr,
StHA~55, and V335 Vul, light-curves were taken from the ASAS database
(Pojma\'nski 2002) \footnote{official home page of ASAS project: {\it
http://www.astrouw.edu.pl/asas/}}. These comprise $V$-band
photometry obtained between November 2000 and February 2009. The ASAS $V$
light-curves are illustrated in Fig.~1, while Table 1 contains the ASAS
names, the mean mags ($\overline{V}$) and the full range of the variations
found in $V$ ($\Delta V$). Although the ASAS database also contains SS73~38,
there are only a few observations of it.

The OGLE-II database (Udalski \etal 1997)\footnote{official home page of OGLE
project: {\it http://ogle.astrouw.edu.pl/}} includes a light-curve
for H2-38. This comprises $I$-band photometry obtained between 1997 and
2000. The OGLE {\it I} light-curve of H2-38 is shown in the second panel of
Fig.~2, while Table 1 gives the OGLE name, the mean mag ($\overline{I}$) and
the full range of the variations found in $I$ ($\Delta I$).

The MACHO database (Alcock \etal 1992) \footnote{official home page of MACHO
project: {\it http://wwwmacho.mcmaster.ca/}} contains observations for
 H2-38, obtained between 1993 and 1999. The photometry was made
through non-standard blue ($B_{\rm M}$) and red ($R_{\rm M}$) filters. The
MACHO light-curve is shown in Fig.~2, while Table 1 gives the MACHO name,
the mean mags ($\overline{B}_{\rm M}$, $\overline{R}_{\rm M}$) and the full
range of the variations found in each band ($\Delta {B}_{\rm M}$, $\Delta
{R}_{\rm M}$).

We also analyse near-IR {\it JHKL} photometry for $o$~Cet, RX~Pup,
V366~Car, BI~Cru, V347~Nor, SS73~38, AS~210, AS~245, RR~Tel and R~Aqr. This
was obtained with the 0.75-m, 1.0-m and 1.9-m telescopes at SAAO and
is on the system described by Carter (1990). The photometry is illustrated
in Figs.~3-5 and mean magnitudes, the full range of variations and the 
amplitudes of pulsation estimated by fitting sinusoids to detrended
 light curves (see Section 3.1..1 for period derivation) are listed in Table 2.
Some of the early data were published and discussed by Feast
\etal 1983a, 1983b; Whitelock \etal 1983; Miko{\l}ajewska \etal 1999;
Kotnik-Karuza \etal 2006, Santander-Garcia \etal 2007.
The near-IR magnitudes of these objects are listed in appendix table and 
available in electronic form
at Acta Astronomica Archive \footnote{official home page of Acta Astronomica: 
{\it http://acta.astrouw.edu.pl/}}.
 
Visual light-curves for $o$~Cet, RX~Pup, BI~Cru, AS~210, RR~Tel and R~Aqr
were extracted from the database of the American Association of Variable
Star Observers\footnote{official home page of AAVSO: {\it
http://www.aavso.org/}}, and used for comparison with the near-IR data.

Most of objects discussed here are definitely symbiotic Miras, with
four exceptions; the symbiotic nature of $o$ Cet, StHA~55 and 
V335~Vul is not certain. 
There have been suggestions (Jura \& Helfand 1984, Kastner \& Soker 2004, Ireland \etal 2007)  that Mira B may be a low mass main-sequence (MS) star. 
StHA~55 and V335~Vul are only suspected of being symbiotic (Belczy{\'n}ski \etal 2000) and may well be normal C-Miras (see also Munari \etal 2008). 
AS~245 is re-classifed here (section 3.1..2) as S-type symbiotic star. These objects are indicated in figures by different symbols.

\begin{sidewaystable}[hpt]
\caption{D-type symbiotic stars in optical massive surveys: designation,
average magnitudes and full range of the variations ($\Delta$) during the
observation run. The first column lists the identification number of the
symbiotic star from Belczy\'nski \etal (2000).}
\begin{center}
\begin{footnotesize}
\begin{tabular}{cccccccccccc}
\hline\hline
\\
No. & Name     &  Name in Survey         & $\overline{V}$ & $\Delta V$ & $\overline{I}$ & $\Delta I$ & $\overline{B}_{\rm M}$ & $\Delta B_{\rm M}$ & $\overline{R}_{\rm M}$ & $\Delta R_{\rm M}$ \\
    &          &                         &                                 \multicolumn{8}{c}{(mag)} \\
\hline
010 & $o$ Cet   & ASAS 021920-0258.6      &    6.75   &    6.28    &     -     &     -      &           -       &            -       &        -          &          - \\
026 & RX Pup   & ASAS 081412-4142.5      &   12.27   &    1.53    &     -     &     -      &           -       &            -       &        -          &          - \\
030 & V366 Car & ASAS 095443-2745.5      &   13.78   &    1.95    &     -     &     -      &           -       &            -       &        -          &          -  \\
034 & BI Cru   & ASAS 122326-6238.3      &   11.44   &    2.90    &     -     &     -      &           -       &            -       &        -          &          - \\
039 & SS73 38  & ASAS 125126-6460.0      &     -     &     -      &     -     &     -      &           -       &            -       &        -          &          -  \\
069 & AS 210   & ASAS 165120-2600.5      &   12.91   &    2.04    &     -     &     -      &           -       &            -       &        -          &          -  \\
120 & H 2-38   & OGLE 180601.19-281704.2 &     -     &     -      &   12.21   &    1.73    &           -       &            -       &        -          &          -  \\
    &          & MACHO 105.21287.23      &     -     &     -      &     -     &     -      &         -9.61     &        $>$0.33     &     -10.78        &         0.96  \\
175 & RR Tel   & ASAS 200419-5543.6      &   11.77   &    0.42    &     -     &     -      &           -       &            -       &        -          &          - \\
188 & R Aqr    & ASAS 234349-1517.1      &    8.40   &    5.86    &     -     &     -      &           -       &            -       &        -          &          - \\
s03 & StHA 55  & ASAS 054642+0643.8      &   13.01   &    2.90    &     -     &     -      &           -       &            -       &        -          &          - \\
s26 & V335 Vul & ASAS 192314+2427.7      &   11.80   &    2.50    &     -     &     -      &           -       &            -       &        -          &          - \\
\hline\hline
\end{tabular}
\end{footnotesize}
\end{center}
\label{tab:1}
\begin{footnotesize}
\end{footnotesize}
\end{sidewaystable}

\begin{sidewaystable}[hpt]

\caption{Near-IR SAAO photometry of the symbiotic stars: average magnitudes
in the $JHKL-$bands, full range of the variation ($\Delta$) of the $JHKL$
magnitude and the amplitudes of pulsation estimated by fitting sinusoid with
peak-to-peak amplitudes of $\Delta_PJ$ etc to the observations.
The first column
lists the identification number of the symbiotic star from Belczy\'nski
\etal (2000).}

\begin{center}
\begin{footnotesize}
\begin{tabular}{cccccccccccccc}\hline\hline\\
No. & Name    & $\overline{J}$ & $\overline{H}$ & $\overline{K}$ & $\overline{L}$ & $\Delta J$ & $\Delta H$ & $\Delta K$ & $\Delta L$ & $\Delta_P J$ & $\Delta_P H$ & $\Delta_P K$ & $\Delta_P L$ \\
    &         &                       \multicolumn{12}{c}{(mag)} \\
\hline
010 & $o$~Cet  &   -1.23 & -2.10 & -2.55 & -3.12 & 1.38 & 1.40 & 1.24 & 1.39 & 1.00 & 1.01 & 0.84 & 0.68  \\
026 & RX~Pup  &    5.67  &  4.18 &  2.98 &  2.28 & 3.72 & 3.01 & 2.30 & 1.59 & 1.08 & 1.00 & 0.81 & 0.64  \\
034 & BI~Cru  &    7.62  &  6.14 &  4.84 &  3.30 & 2.07 & 1.40 & 1.11 & 0.93 & 0.69 & 0.46 & 0.31 & 0.19  \\
030 & V366~Car &   7.19  &  5.77 &  4.78 &  3.52 & 3.05 & 2.58 & 1.93 & 1.28 & 0.53 & 0.59 & 0.49 & 0.39  \\
039 & SS73~38 &    9.41  &  7.47 &  5.96 &  4.18 & 3.18 & 2.56 & 1.79 & 1.06 & 0.53 & 0.59 & 0.49 & 0.39  \\
060 & V347~Nor &   7.01  &  5.54 &  4.78 &  4.02 & 1.81 & 1.32 & 1.00 & 0.94 & 0.88 & 0.70 & 0.53 & 0.44  \\
069 & AS~210  &    9.50  &  7.74 &  6.28 &  4.60 & 2.67 & 2.48 & 1.84 & 1.27 & 1.06 & 1.16 & 1.05 & 0.85  \\
105 & AS~245  &    9.41  &  8.05 &  7.44 &  6.92 & 0.54 & 0.57 & 0.53 & 0.44 & 0.36 & 0.42 & 0.38 & 0.30  \\
175 & RR~Tel  &    6.57  &  5.37 &  4.43 &  3.13 & 3.47 & 2.84 & 2.05 & 1.29 & 0.71 & 0.69 & 0.60 & 0.47  \\
188 & R~Aqr   &    0.80  & -0.27 & -0.89 & -1.72 & 3.58 & 3.07 & 2.42 & 1.79 & 1.00 & 0.91 & 0.78 & 0.72  \\
\hline\hline
\end{tabular}
\end{footnotesize}
\end{center}
\label{tab:2}
\end{sidewaystable}

\begin{table}[hpt]
\caption{Pulsation periods derived from our analysis.}
\begin{center}
\begin{footnotesize}
\begin{tabular}{cccccccc}\hline\hline\\
No. & Name    & Type &    ASAS      &    OGLE      &    MACHO     &    SAAO     & Other periods \\
    &         &      &           \multicolumn{5}{c}{(days)}   \\
\hline
010 & $o$~Cet    &  O  &  338$\pm$10  &      -       &      -       &  332$\pm$3  & 331.96$^1$ \\
026 & RX~Pup   &  O  &    trend     &      -       &      -       &  575$\pm$8  & 578$^2$ \\
030 & V366~Car &  O  &    trend     &      -       &      -       &  432$\pm$6  & 433$^3$ \\
034 & BI~Cru   &  O  &  280$\pm$7   &      -       &      -       &  274$\pm$2  & 280$^4$\\
039 & SS73~38  &  C  &      -       &      -       &      -       &  463$\pm$7  & \\
060 & V347~Nor &  O  &      -       &      -       &      -       &  374$\pm$4  & 373$^5$ \\
069 & AS~210   &  C  &  407$\pm$14  &      -       &      -       &  423$\pm$7  & \\
105 & AS~245   &  O  &      -       &      -       &      -       &  366?       & \\
120 & H~2-38   &  O  &      -       &  425$\pm$36  &  395$\pm$16$^{\rm a}$  &      -      & \\
175 & RR~Tel   &  O  &    trend     &      -       &      -       &  385$\pm$4  & 385$^6$ \\
188 & R~Aqr    &  O  &  395$\pm$13  &      -       &      -       &  391$\pm$5  & 386.96$^1$ \\
s03 & StHA~55  &  C  &  372$\pm$15  &      -       &      -       &      -      & 395$^7$\\
s26 & V335~Vul &  C  &  334$\pm$14  &      -       &      -       &      -      & 347$^8$ \\
\hline\hline
\end{tabular}
\end{footnotesize}
\end{center}
\label{tab:2}
\begin{footnotesize}
$^{\rm a}$ derived from $R_{\rm M}$.\\ 
References: [1] Kholopov 1985;
[2] Miko{\l}ajewska \etal 1999; [3] Feast \etal 1983b;
[4] Whitelock \etal 1983; [5] Santander-Garcia \etal 2007;
[6] Kotnik-Karuza \etal 2006; [7] Munari \etal 2008; [8] Sobotka \etal 2003.
\end{footnotesize}
\end{table}

\begin{table}[hpt]
\caption{New pulsation ephemerides.}
\begin{center}
\begin{footnotesize}
\begin{tabular}{ccl}\hline\hline\\
No. & Name    & Ephemeris  \\
&&\\
\hline

026 & RX~Pup   &  ${\rm Max}(JHKL)=2\,442\,238 + 575\times E$ \\
030 & V366~Car &  ${\rm Max}(JHKL)=2\,442\,413 + 432\times E$ \\
034 & BI~Cru   &  ${\rm Max}(JHKL)=2\,443\,291 + 274\times E$ \\
039 & SS73~38  &  ${\rm Max}(JHKL)=2\,444\,983 + 463\times E$ \\
069 & AS~210   &  ${\rm Max}(JHKL)=2\,446\,162 + 423\times E$ \\
120 & H~2-38   &  ${\rm Max}(R_{\rm M})=2\,449\,205 + 395\times E$ \\
175 & RR~Tel     &  ${\rm Max}(JHKL)=2\,442\,207 + 385\times E$ \\
s03 & StHA~55  &  ${\rm Max}(V)=2\,452\,2712 + 372\times E$ \\
\hline\hline
\end{tabular}
\end{footnotesize}
\end{center}
\label{tab:eph}
\end{table}

\Section{Analysis and Results}

\subsection{Variability}

\subsubsection{Period analysis}

All light-curves were analysed using the program PERIOD\footnote{the source
program is available on {\it http://www.starlink.rl.ac.uk/}} ver. 5.0, based
on the modified Lomb-Scargle method (Press \& Rybicki, 1989). Long-term
trends were removed by subtracting a polynomial of appropriate order, and the
resultant power spectra were compared with the window spectra.  The periods
were derived from the inverse of the maximum of the peak in the periodogram
($P=f_{max}^{-1}$), whereas their accuracy was estimated by calculating the
half-size of a single frequency bin ($\Delta f$), centred on the peak
($f_{c}$ is the center of the peak ) of the periodogram and then 
converted to period units ($\Delta P =f_{c}^{-2} \cdot \Delta f$). 
The results of our period analysis are summarized in Table 3.  
Examples of our power spectra are shown in Fig.~6-7 and Figs.~8 shows
near-IR light curves folded with pulsation periods.

The highest peak in a typical power spectrum corresponds to the pulsation
period, while the other peaks represent annual aliases, second and third
harmonics, long-term variation and some combination thereof. The only
exceptions are the ASAS light-curve of V335~Vul, where the highest peak
corresponds to the second harmonic, and the near-IR light-curves of RR~Tel
where it represents the annual alias. In both exceptional cases this is due
to gaps in the light-curve of the object.

There is practically no difference between the power spectra derived for the
$JHK$ or $L$ observations.  They all show the same position of peaks with a
little difference in power. The latter is due to differences of amplitudes of
the pulsations and long-term trends in the near-IR photometry.

For four systems (AS~210, H2-38,  SS73~38, and AS 245) the
pulsation periods are derived for the first time.  Pulsations are also
detected in other systems with known periods. In particular, pulsations are
always visible in the near-IR light-curves of all of the symbiotics we
examined. However, the periods for $o$~Cet and R~Aqr (Table 3) are not as
accurate as those derived from visual data collected over a few centuries
(Kholopov 1985), whereas for RX~Pup, V366~Car, RR~Tel and BI~Cru our new
estimates are better than published values (Miko{\l}ajewska \etal 1999,
Feast \etal 1983a, Whitelock \etal 1983) because they are based on more
data. For some systems pulsations are also visible in the optical
light-curves, although the resulting periods are less accurate than those
already known (Belczy{\'n}ski \etal 2000, and references therein) because
ASAS, MACHO and OGLE cover relatively short time periods (a few to several
years). Finally, the optical light-curves of RX~Pup, V336~Car and RR~Tel do
not show pulsations presumably because the Mira in these systems is
obscured by optically thick dust and/or the optical light is dominated by
emission from the hot component.

The near-IR measurements for AS~245 suggest significant variability and a
period of the order of one year, which we adopt for the following
discussion. However, with only 12 observations spread over almost two
decades, the data are inadequate for a proper analysis.

Recently, Munari \etal (2008) estimated a pulsation period of 395 days for
StHA~55 in the $V$-band. The value is very uncertain because they had only observations covering only one pulsation cycle. The pulsation period of StHA~55 derived from ASAS data, which covers 6 pulsation cycles, is 372 days. 

The pulsations ephemerides for three systems for which
periods are derived here for the first time (AS~210, H2-38, SS73~38) and for StHA~55, RX~Pup, V366~Car, RR~Tel, and BI~Cru are listed in Table 4.

Fig.~9 presents the distribution of the pulsation periods for the symbiotic
Miras together with those for non-symbiotic Miras. For the symbiotic Miras
we have combined the published periods (Belczy{\'n}ski \etal 2000) with our
new estimates (Table 3). The data for the non-symbiotic Miras are from
Olivier \etal (2001), and Whitelock \etal (1994, 2000, 2006). These
non-symbiotic Miras are henceforth referred to as `normal Miras', but note
that they will include widely separated binary systems; indeed $o$ Ceti
itself is included in both the symbiotic and normal groupings. The normal
Miras show distinct period distributions, with peaks at $\sim 330$ and $\sim
530$ days for the O- and C-rich objects, respectively. The period
distribution for normal Miras is influenced by selection effect that are
extremely difficult to quantify. We do know that there are O-rich Miras with
periods over 1000 days, the OH/IR stars, but these have progenitors of
several solar masses and are quite rare. The symbiotic Miras have a mean
period of about 400 days, and with one exception range from 280 (BI Cru) to
580 (RX Pup). The exception, V407 Cyg, has the period of 763 days (Kolotilov
\etal 2003) and is thought to be hot bottom burning giant (Tatarnikova \etal 2003)
and is therefore probably more massive than its shorter period counterparts.
There is no obvious difference between the periods of C-rich (5 objects) and
O-rich (19 objects) symbiotic Miras. All O-rich symbiotic Miras have periods
longer than the median value for their normal counterparts. Whitelock (1987)
suggested that this was essentially a selection effect - mass transfer in
these systems is via the stellar wind of the Mira and long period Miras have
in general higher mass-loss rates. Thus symbiotic activity will be seen
from white dwarfs in binary systems with longer period Miras at much larger
separations than from those in binaries involving shorter period Miras, {\it i.e.}
we expect there to be unidentified white dwarfs in binary systems with short
period Miras with very low levels of interaction.

\subsubsection{Long term trends}

The near-IR light-curves of all symbiotic Miras included in this study show,
in addition to the Mira pulsations, significant long-term variations. Such
trends are very common in symbiotic Miras, and are almost certainly caused
by variable dust obscuration ({\it e.g.} Whitelock 1987; Miko{\l}ajewska \etal
1999).

To get better insight into this phenomenon, we removed the Mira pulsations
from the light-curves and examine the trends in the residuals. In the case
of stars with periods longer than 400 days this is done by subtracting the
best fitting sine-curve. For shorter period targets, $o$~Cet, R~Aqr, BI~Cru
and V347~Nor, the pulsation light-curves are asymmetrical and simply
subtracting the sine-curve produced a lot of scatter in the residual. 
Therefore another method was used. We first generated an average pulsation
curve and then subtracted it from the original light-curve. This method
produced less scatter for the short period stars, while for the long period
pulsators both methods give similar results. In the case of BI~Cru, data
between JD2\,445\,400 and JD2\,451\,400 were used to prepare the average
pulsation light-curves by excluding the dust obscuration events. In the
case of R~Aqr, data before JD2\,444\,000 were omitted from the average
pulsation light-curves for the some reason. 

The light-curves, prior to removal of the pulsations, are plotted in the
upper panels of Figs.~3 to 5. The secular trend is clearest at $J$ and the
behaviour of $J-K$ indicates that it is due to increasing reddening, as
expected for increasing optical depth of the dust. This general trend is
also present in the $L$ light-curves.

The near-IR colours of D-type systems indicate the presence of warm dust
shells. {bf Fig.~10} presents the $J-K$ vs. $K-L$ diagram for the symbiotic Miras
in this study together with those for normal Miras. The symbiotic
Miras are shown in both their obscured and unobscured states. This
demonstrates that most of the colours can be qualitatively reproduced with a
shell of around 800K (see also Whitelock 1987). In at least 50 \% of the
studied systems the reddening toward the Mira star was larger than reddening 
toward the hot component ({\it e.g.} Miko{\l}ajewska
1999). There appears to be very much less extinction towards the high
excitation regions (emission lines and hot UV continuum) than towards the
Mira suggesting that the hot component generally lies outside the dust
cocoon associated with the Mira. This fact constrains the orbital
separations in these systems to be $\gtrsim$ 10-15 AU, and the orbital
period to be $\gtrsim$ 20 yr, because the typical radius of a dust shell is
$\gtrsim$ 5 $R_{Mira}$, and the Mira radius, $R_{Mira} \sim$ 2-3 AU 
(Miko{\l}ajewska 1999).

In the specific case of R Aqr the obscuration was explained in terms of
orbitally related eclipses of the Mira by the accreting stream (Gromadzki \&
Miko{\l}ajewska 2009, Willson, Garnavich \& Mattei 1981). However, the dust
obscuration phenomenon in symbiotic Miras cannot be, in general, orbitally
related (Miko{\l}ajewska 1999). First, these events seem to occur in well
observed symbiotic Miras with too great a frequency, and in several systems
with more than one event observed, they were separated by $\sim 2000-4000$
days, much below the minimum orbital period expected for these systems.
Secondly, while in RX~Pup the infrared emission from the Mira shows
modulation with a time scale of $\sim 3000$ days with two apparent
minima around 1984 and 1990, at the same time the reddening towards the hot
component was constant over at least 13 years (Miko{\l}ajewska \etal
1999). A geometry of the binary system in which the cool component is
eclipsed at least twice, while the hot one is not eclipsed at all is
impossible. Moreover, the fact that no D-type system exhibits greater
extinction towards the hot component than towards the Mira implies that only
the Mira is affected. All these point to intrinsic variations in the Mira
envelope as the main cause of the dust obscuration (see Sec. 3.2.2). 

For most of these systems the obscuration of light at $J$ is not associated
with a brightening at $L$, as might be anticipated if the dust forms in a
uniform shell around the Mira (see Figs.~3 to 5).  BI~Cru is the one
exception, but it has a spectrum quite unlike that of a normal
O-rich Mira in that it does not show the characteristic H$_2$O absorption. 
What it does show is very strong 2.3 $\mu$m CO-band emission, and in that
respect it has similarities to certain B[e] stars, such as Hen~3-1138 and
Hen~3-1359, (Whitelock \etal 1983). It is therefore possible that we are observing a slightly different phenomenon in this case. We note that all parameters derived below (distance, mass-loss rate etc) for BI~Cru through relations for normal Miras must be regarded as uncertain.

C-rich Miras, including R~For,
also show obscuration events that cannot be reconciled with spherically
symmetric dust ejection (Whitelock \etal 1997; Feast \etal. 2003; Whitelock
\etal 2006). The alternative scenarios involve ejection around an equatorial
disk or as puffs in random directions. In the cases of RR~Tel and AS~210
spectro-polarimetry suggests a low inclination orbit (Schmid \& Schild
2002), but they show strong dust obscuration. This fact rather favours the
ejection of puffs in random directions as was suggested for C-Miras (Feast
\etal 2003; Whitelock \etal 2006) and is well known among R CrB stars.

Secular trends are visible in the optical (ASAS) light-curves of RX~Pup,
V366~Car and RR~Tel. These stars do not show pulsational variations in the
visible and the trends observed are not well correlated with the long term
changes in their near-IR light-curves. RR~Tel and RX~Pup underwent nova
outbursts in the 1940s and 1970s, respectively, and the nearly linear trends
in their optical light-curves are due to the fading of their hot components
and nebular radiation.  The cause of the optical trend in V366~Car is less
clear, although the presence of a minimum with a time scale similar to the
dust obscuration observed in its near-IR light-curve at earlier epochs may
indicate some connection with a dust obscuration event.

The visual light-curve of $o$~Cet  shows
cycle-to-cycle changes of the visual amplitude. Similar behaviour is seen in
most Miras with sufficiently good light-curve coverage. Erratic variations
are also evident in the near-IR data, but the coverage of the light-curves
is not as good. We note, however, that the highest visual maxima are
correlated with the maxima in near-IR light-curve with (see left panels of Fig. 3) which
may suggest changes of the average luminosity of the Mira.

AS~245 has colours typical of an oxygen-rich Mira (see Fig.~10).
However, the amplitude of the possible pulsation ($\lesssim0.4$ mag in all
bands) is lower than in other symbiotic, and even non-symbiotic, Miras. Low
amplitude and uncertain pulsation suggest that the cool component of AS~245
is an SRa variable rather than the Mira.
We also point out similarity between the cool giant of AS~245 and that in
S-type symbiotic MWC~560 (V694~Mon). The latter also shows low-amplitude
pulsations with a period of $\sim$340$^{\rm d}$ and low $K-[12]$. Gromadzki \etal
(2007) suggested that the red component of MWC~560 is on thermally-pulsating
AGB, although its pulsation characteristics may be influenced by its nearby
companion possibly reducing the pulsation amplitude and the circumstellar
dust. More observations are necessary to
settle the nature of the cool component of AS~245, but we have enough
information here to reclassify it as an S-type; its position in Fig.~10
does not show signs of the dust associated with most symbiotic Miras.

\subsection{Physical parameters}

The great advantage of long-period pulsating AGB stars is an opportunity to
determine various physical parameters such as absolute magnitudes, distances
and mass-loss rates using near-IR period-luminosity-colour relations derived
by Feast, Whitelock and coworkers in a series of papers. In this section we
apply these methods to estimate the physical parameters for symbiotic
systems containing C-rich and O-rich Miras.

\subsubsection{Extinction and distance}

The absolute $K$ magnitudes of both O- and C-rich symbiotic Miras are
estimated using the latest Whitelock \etal (2008) period-luminosity (PL)
relationship:

\begin{equation}
M_{K} = \rho[\log P-2.38]+\delta .
\end{equation}

The slope $\rho=-3.51\pm0.20$  was derived from large amplitude asymptotic
giant branch variable in the LMC.  We use the zero-point
$\delta=-7.25\pm0.07$ for O-rich Galactic Miras, estimated using the revised
{\it Hipparcos} parallaxes together with published VLBI parallaxes for OH
Masers and Globular Clusters. That value agrees with those estimated for
O-rich LMC Miras ($\delta=-7.15\pm0.06$) and C-rich Galactic
($\delta=-7.18\pm0.37$) and LMC ($\delta=-7.24\pm0.07$) Miras (assumes an LMC
distance modulus of $18.39\pm0.05$ mag; van Leeuwen \etal 2007).

The interstellar extinction at $V$, $A_V^{\rm D}$, is estimated using the
Drimmel \etal (2003) 3-D Galactic extinction model, including the rescaling
factors that correct the dust column density to account for small structures
observed in the DIRBE data, but not included explicitly by the model. The
extinction at $JHK$ is then calculated from the relations given by Glass
(1999) for photometry on the SAAO system. Obviously this extinction
represents only the interstellar component and tells us nothing about any
circumstellar reddening.

We can compare the observed $J-K$ with the intrinsic value, as discussed
below, to get an estimate of the total, interstellar 
plus circumstellar extinction.
The intrinsic $(J-K)_0$ for O Miras is estimated using
the  period-colour relation derived by Whitelock
\etal (2000) for Miras in the solar neighbourhood observed by Hipparcos:

\begin{equation}
(J-K)_0 = 0.71 \log P-0.39 .
\end{equation}

Intrinsic near-IR colours $(J-K)_0$ of C Miras are obtained from the
period-colour relations for C Miras from Whitelock \etal(2006):
\begin{equation}
(J-K)_0 = 12.811 \log P -30.801 .
\end{equation}

Although the period colour relation for O-Miras is quite well defined, at
least at short periods, that for C-rich Miras is not; for example
the best estimated relation, $(H-K)_0$ vs. period, has a standard deviation of
$\sigma=0.48$ mag. Therefore, a reliable total extinction estimate is
often impossible for C-rich objects and the uncertainty of the values
derived for $E(B-V)$ are huge.

We assume wherever possible that the extinction derived from the
period-colour relations is the total, circumstellar plus interstellar,
value and that subtracting the interstellar extinction, evaluated as
described above, gives the circumstellar value.
The values of $E(B-V)$ in the unobscured parts of light-curves, that are
used for these estimates, are listed in Table 5.

To estimate distances we derive the absolute $K$ mag, $M_K$, from
the period and take the observed $K$ mag during intervals in which the
objects did not show obvious dust obscuration, and average over the
pulsation cycle.  We use $A_K$ derived from the colour excess to correct for
the interstellar plus circumstellar extinction. The formal error of the
distance derived by this approach is 12 -- 20 \%. The parameters derived as
described above are listed in Table~4.

\subsubsection{Mass loss}

There is a good correlation between mass-loss rate and $K-[12]$ colour for
both C- and O-rich Miras, provided we can assume that the shells are
approximately spherically symmetric. This arises because to a first
approximation the $K$ and [12] mags are measures of the brightness of the
star and of the shell, respectively. If, however, the shell is very
asymmetric and our line of site to the star is obscured by thick dust (as it
would be during an obscuration event) then we will underestimate the $K$
brightness and overestimate the mass-loss rate. With this caveat in mind we
can estimate the mass-loss rates for the symbiotic Miras from their $K-[12]$
colours (Table 5).

In the case of carbon Miras, Whitelock \etal (2006) estimated mass-loss
rates using a modification of Jura's (1987) method, and fitted an analytical
formula which we use here:
\begin{eqnarray}
\log \dot{M}&=& -7.668 + 0.7305 (K-[12]) - 5.398 \times 10^{-2} (K-[12])^2 \\ 
&& + 1.343 \times 10^{-3} (K-[12])^3. \nonumber
\end{eqnarray}

The mass-loss rates for the symbiotic O-rich Miras are determined using
the correlation with $K-[12]$ colour derived for 58 high mass-loss O-rich AGB
stars in the South Galactic Cap (Fig.~21 of Whitelock \etal 1994).

The $[12]$ magnitudes are calculated from the {\it IRAS} 12-micron fluxes
(Belczy\'nski \etal 2000) using $[12]=-2.5 {\rm log} F_{12} + 3.62$.  The
$K$ magnitudes for the objects with SAAO near-IR light-curves are the values
observed outside of obvious obscuration phases and averaged over the
pulsation cycle. For HM~Sge and V1016~Cyg we derived the average $K$ from their published near-IR light-curves (Taranova \& Shenavrin 2000).  The use of a $K$ magnitude obtained during a dust obscuration phase could result in an overestimate of the mass-loss rate by an order of magnitude (we assume, as
suggested above, that the faint phases are the result of asymmetric
obscuration and that the [12] mag remains approximately constant). For
example the average $K$ mag for V366~Car outside dust obscuration is 4.4
mag, from which we derive a mass-loss rate of $6.3 \times 10^{-7} M_{\odot}
{\rm yr}^{-1}$; during dust obscuration $K$ is about 6 mag and the derived
mass-loss rate is $5 \times 10^{-6} M_{\odot} {\rm yr}^{-1}$. The difference is not so large for objects with greater mass-loss rates, {\it e.g.}, RX~Pup
outside of dust obscuration has $\dot{M}=6.3 \times 10^{-6} M_{\odot} {\rm
yr}^{-1}$, which increases to $\dot{M}=10^{-5} M_{\odot} {\rm yr}^{-1}$
during obscuration.

For the remaining objects we use data from  2MASS (obtained
between 1997 and 2000) and Munari \etal (1992; obtained in 1990). Whereas
the near-IR data in Munari \etal (1992) are from SAAO using the same
photometric system as the light-curves presented here, the $K_{\rm s}$
magnitudes from 2MASS have been transformed to the SAAO system using
transformations from Carpenter (2001). A significant fraction of symbiotic
Miras show dust obscuration events, so it is possible that some of these
measurements were made during such events. 
Typically, for objects with SAAO near-IR light-curves, the difference
between the average $K$ and transformed 2MASS $K_s$ is 0.1-0.2 mag, and so
we adopted the average of the Munari \etal (1992) and transformed 2MASS
values. The $K$ magnitudes have been corrected for interstellar reddening
(section 3.2..1).

We should emphasize that mass-loss rates obtained in this way are not
accurate. It is also important to appreciate that we do not yet understand
the obscuration events or their link to the Mira mass-losing activity. If
the symbiotic stars show obscuration events for the same reason that Feast
(2003) and Whitelock \etal (2006) have suggested that the C-Miras do, then
the mass-loss will be in discreet dust puffs in random directions like RCB
stars (see also Whitelock 2003).  
The fact that symbiotic Miras are in binary
systems will almost certainly effect what happens to the dust once it leaves
the immediate vicinity of the Mira in a non-random fashion, even if it does
not affect the dust production itself.

Fig. 11 presents the $K_0-[12]$ colour distribution for symbiotic Miras and
compares it to those for normal Miras. 
The normal O-rich Miras show a distinct $K_0-[12]$ colour distribution,
which peaks at 1.76 mag, corresponding to a mass-loss rate of $\sim 10^{-7}
M_{\odot} {\rm yr}^{-1}$, and has a tail that extends across the region
occupied by the symbiotic Miras. As can be seen from Fig.~12, a plot of the
$K_0-[12]$ colour vs. pulsation period, the tail is comprised of long period
(mostly $P>400$ days) objects. The C-rich Miras show a broad distribution
from 1 to 9.5 mag which implies mass-loss rates ranging from $\sim 10^{-7}$
to $\sim 3\times10^{-5} M_{\odot} {\rm yr}^{-1}$. The symbiotic Miras show a
distinct $K_0-[12]$ colour distribution with a peak at 4.25 mag,
corresponding to a mass-loss rate of $\sim 3.2\times10^{-6} M_{\odot} {\rm
yr}^{-1}$. These figures show that the symbiotic Miras have on average
larger $K_0-[12]$ than the their normal counterparts with similar periods. 
Fig.~13 shows a plot of $K_0-[12]$ vs. $(J-K)_{0}$ for symbiotic and normal
Miras.
The scatter of symbiotic Miras is larger than that of the normal Miras
and, although the distributions overlap, the symbiotic Miras have larger
$K-[12]$ than normal Miras of the same period or same $J-K$. If this is the
result of selection effects then we are only finding symbiotic Miras at the
very end of their evolution when mass-loss is at its maximum. It seems more
likely to be a consequence of the binary interaction.

The widely accepted mass-loss scenario ({\it e.g.} Fleischer, Gauger \&
Sedlmayer 1992) is that the Mira pulsation lifts matter above the
atmosphere, but does not accelerate it to escape velocity. As the matter
falls it encounters the next pulsation, which gives it another outward
impulse. Matter is pushed by pulsations until its temperature drops enough
for dust to condense (at a few stellar radii). Then, radiation pressure on
the dust can efficiently accelerate the dust, and the dynamically coupled
gas, to the escape velocity. $K_0-[12]$ is proportional to the amount of
dust between us and the Mira, as described above. We should emphasize that in
a binary system higher $K_0-[12]$ could be the result of the lost mass not
leaving the system rather than of intrinsically higher mass-loss rates.  
For example, if mass lost from the Mira in the binary was trapped within 
the system ({\it e.g.} near the $L_{4}$ and $L_{5}$ Lagrange points), more 
cool dust in the system will mean more obscuration and more flux at 
12~$\mu$m and will result in higher values of $K_0-[12]$.

An important question is about the influence of the secondary star on the
mass-loss rate and its character. Podsiadlowski \& Mohamed (2007) proposed a
wind Roche lobe overflow (RLOF) model for $o$ Cet. In their model, a slow
wind from Mira fills its Roche lobe and then the matter streams -- via the
L1 Lagrangian point -- onto an accretion disk around the companion. This
models works most efficiently when the radius of dust shell is comparable
with the Roche lobe radius ($R_{\rm RL}$).  Orbital separations for
symbiotic Miras are essentially unknown, with the exception of R~Aqr. The
mean angular rotation rate from spectro-polarimetry suggest an average
period of around 150 yr (Schmid \& Schild 2002). This estimation is
uncertain because their observations cover about 10 yr (only about 5\% of
the implied orbital period). For typical component masses, $M_{\rm
hot}\sim0.6 {\rm M_{\odot}}$ and $M_{\rm Mira}\sim1.2 {\rm M_{\odot}}$, this
period corresponds to an orbital separation of $\sim 40$ AU and $R_{\rm RL}
\sim 15$ AU. Thus wind RLOF should occur in most symbiotic Miras.
Podsiadlowski \& Mohamed (2007) made SPH simulations, which show that in a
case similar to symbiotic Miras most of the material from the wind should be
accreted or remain bound to the system, and only very small fraction is
ejected to infinity, {\it i.e.}, the accretion rate is $\sim$100 times
higher than the Bondi-Hoyle value and the mass loss will tend to be confined
to the orbital plane.


\begin{sidewaystable}[hpt]

\caption{The physical parameters of the symbiotic stars discussed here:
distance ($d$), mean magnitude in the $K-$band outside of dust obscuration
phases, absolute magnitude in $K-$band ($M_{K}$), average $J-K$ colour
($\overline{J-K}$), unredded $J-K$ colour derived form colour-period
relation ($(J-K)_{0}$), interstellar
reddening estimated from a 3-D extinction model of Galaxy ($E_{B-V}^{\rm
D}$), total reddening estimated from colour excess ($E_{B-V}$), $K_{0}$-[12]
colour and logarithm of mass-loss rate ($\log \dot{M}$). }

\begin{center}
\begin{footnotesize}
\begin{tabular}{ccccccccccc}\hline\hline
\\
No. & Name     & $d$  & $K$   & $M_{\rm K}$ & $\overline{J-K}$ & $(J-K)_{0}$ & $E_{B-V}^{\rm D}$ & $E_{B-V}$ & $K_{0}$-[12] & $\log{\dot{M}}$ \\
    &          & (kpc) &         \multicolumn{7}{c}{(mag)}                                              & ($M_{\odot} {\rm yr}^{-1}$) \\
\hline
010 & $o$ Cet    & 0.11 & -2.55 &   -7.75   &      1.32        &   1.40    &       0.01    &   -   &   3.04   &    -6.2 \\
026 & RX Pup   & 1.6 &  2.80 &   -8.58   &      2.36        &   1.57    &       0.51    &  1.51 &   4.90   &     -5.2 \\
030 & V366 Car & 2.8 &  4.40 &   -8.15   &      2.07        &   1.48    &       0.59    &  1.13 &   3.06   &     -6.2 \\
034 & BI Cru   & 2.3 &  5.02 &   -7.49   &      2.58        &   1.34    &       0.91    &  2.36 &   4.25   &     -5.5 \\
039 & SS73 38  & 4.8 &  5.34 &   -8.27   &      3.09        &   3.40    &       0.76    &   -   &   3.68   &     -5.6\\
060 & V347 Nor & 2.9 &  4.78 &   -7.93   &      2.23        &   1.44    &       0.49    &  1.51 &   2.83   &     -6.3 \\  
069 & AS 210   & 5.6 &  5.62 &   -8.11   &      2.72        &   2.85    &       0.27    &   -   &   3.43   &     -5.7\\ 
105 & AS 245   & 10.2 &  7.44 &   -7.89   &      1.99        &   1.43    &       1.32    &  1.03 &   4.01   &    -5.6 \\
120 & H 2-38   & 7.2 &  6.60 &   -8.01   &      1.90        &   1.45    &       0.82    &  0.85 &   3.91   &     -5.7 \\ 
175 & RR Tel   & 2.5 &  4.18 &   -7.97   &      1.71        &   1.45    &       0.04    &  0.51 &   3.79   &     -5.8 \\
188 & R Aqr    & 0.24 & -1.06 &   -7.99   &      1.59        &   1.45    &       0.02    &  0.23 &   3.38   &    -6.0 \\
s03 & StHA 55  & 3.6 &  5.27 &   -7.92   &      2.99        &   2.13    &       0.23    &  1.64 &   2.58   &     -6.1\\
s26 & V335 Vul & 3.7 &  5.11 &   -7.87   &      1.85        &   1.95    &       0.50    &   -   &   1.97   &     -6.4\\
\hline\hline
\end{tabular}
\end{footnotesize}
\end{center}
\label{tab:4}
\begin{footnotesize}
\end{footnotesize}
\end{sidewaystable}

\Section{Conclusions}

The study of symbiotic Miras is a difficult task because of the presence of an active accreting companion and the long time-scales of the variations (pulsation $\sim$1-2 years, dust obscuration $\sim$10 years, orbital period $\sim$100 yr). Fortunately data and relations estimated for normal Miras allowed us to derive some relevant physical  
parameters, such as distances, luminosities and order of mass-loss rates
(provided certain assumptions are made).      

A comparison of the symbiotic Miras with normal Miras of similar           
pulsation period shows that the symbiotic stars have on average higher
values of $K_{0}-[12]$. This may indicate that they have higher mass-loss
rates, or more likely that the dust which is being lost by the Miras is
trapped within the binary system. Our observations do not settle this issue.    
Undoubtedly, mass-loss rates and mass transfer in these systems are most interesting
subjects, but more observational data, especially in UV, IR and radio bands, are needed 
to understand it.

\Acknow{We are grateful to the following people who made observations from
SAAO that were included in this analysis: Greg Roberts, Robin Catchpole, Brian Carter, Dave 
Laney and Hartmut Winkler.
We also would like to thank Magda Borawska and Anna Lednicka for their work
during initial phase of this project. This study made use of the American
Association of Variable Star Observer (AAVSO) International Database
contributed by observers worldwide and the public domain databases of The
We are grateful to Michael Feast for his comments on an early draft of this
paper.
All Sky Automated Survey (ASAS) and The Optical Gravitational Lensing
Experiment (OGLE), The Two Micron Sky Survey (2MASS), The MACHO Project
(MACHO) which we acknowledged. This research has made use of the SIMBAD
database, operated at CDS, Strasbourg, France. This work was partly
supported by the Polish Research Grants No. 1P03D\,017 27 and N203\,395534.}

\newpage
\begin{figure}
\includegraphics[angle=0,width=13cm]{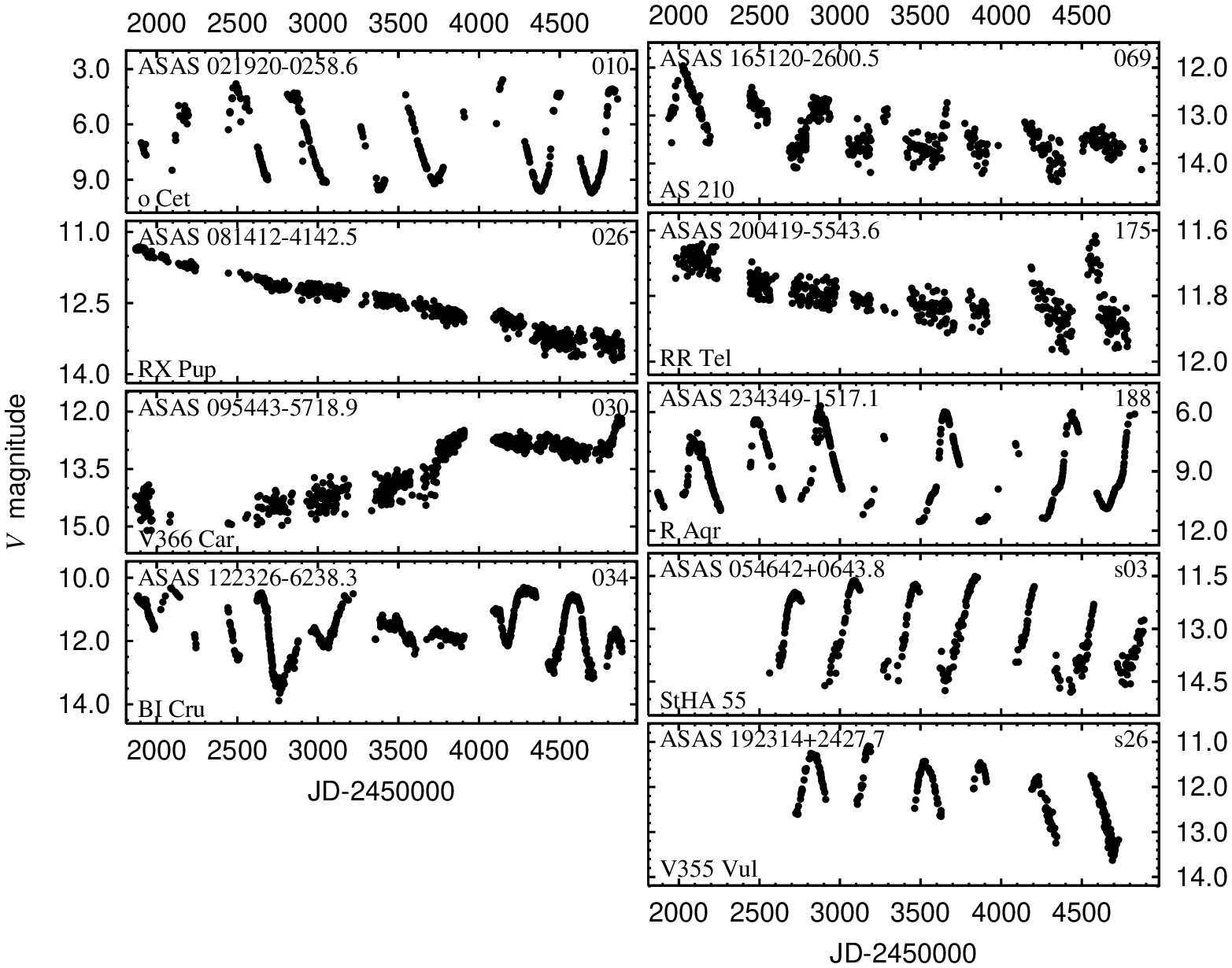}
\caption{ASAS light curves of type-D symbiotic stars.}
\end{figure}

\newpage
\begin{figure}
\begin{center}
\includegraphics[width=11.5cm]{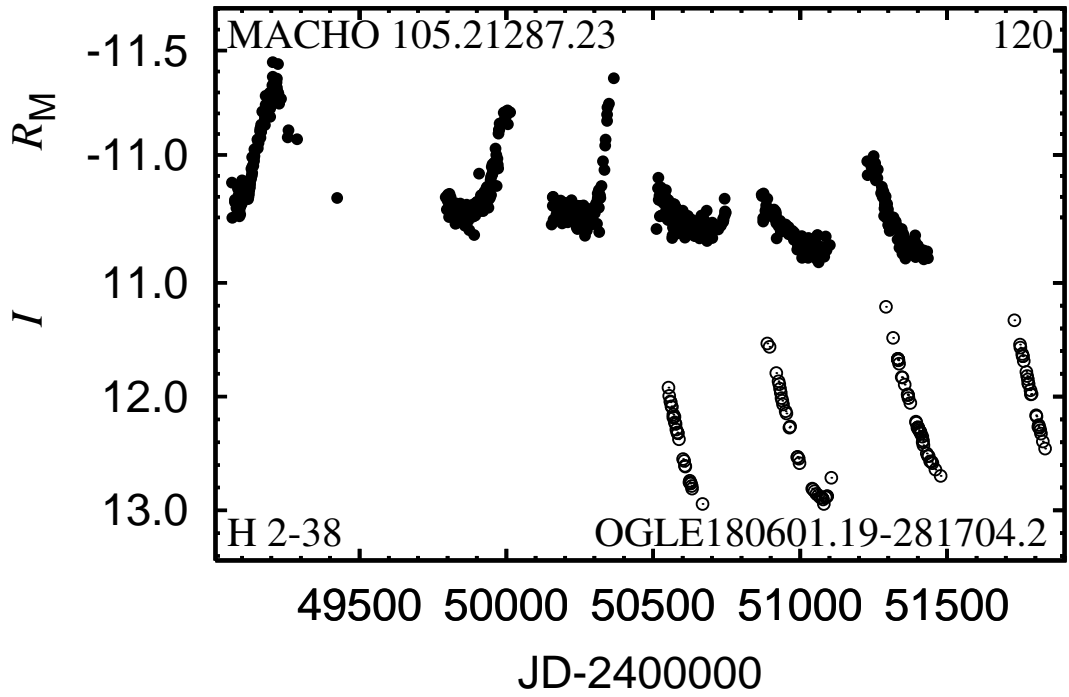}
\end{center}
\caption{MACHO (filled circle) and OGLE (open circle) light curve of H~2-38.}
\end{figure}

\clearpage
\begin{sidewaysfigure}
\includegraphics[width=18cm]{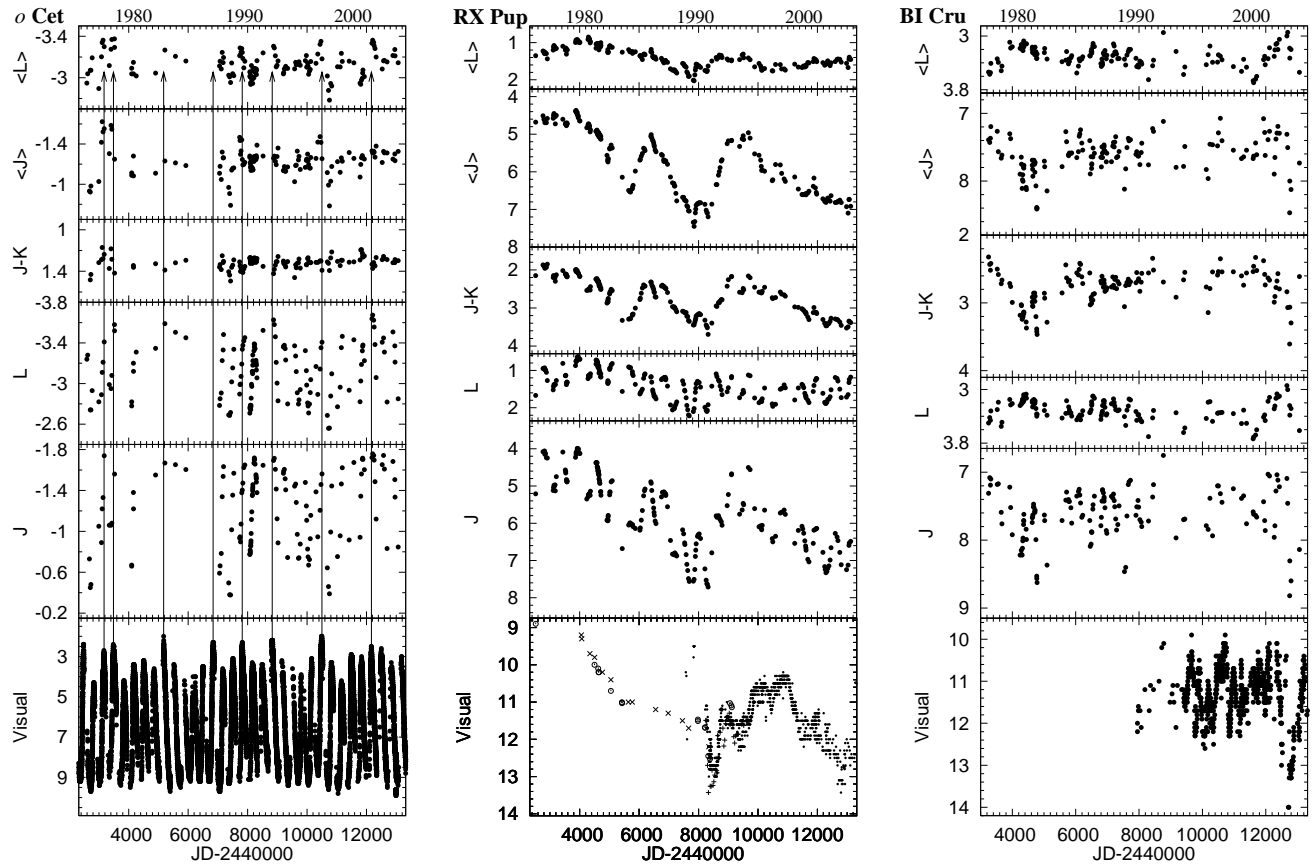}
\caption{Near-IR light curves of $o$ Cet (left panels), RX Pup (middle panels) and BI Cru (right panels). 
In left panels arrows mark maxima.}
\end{sidewaysfigure}

\clearpage
\begin{sidewaysfigure}
\includegraphics[width=18cm]{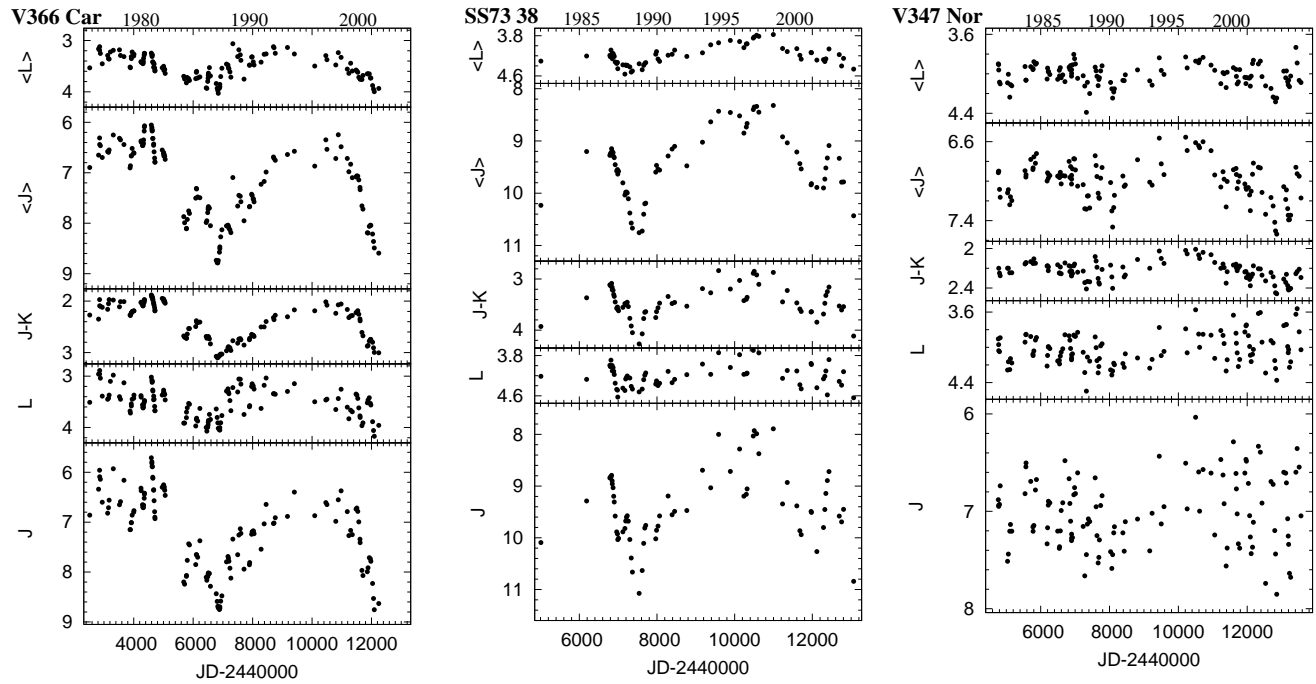}
\caption{Near-IR light curves of V366 Car (left panels), SS73 38 (middle panels) and V347 Nor (right panels).}
\end{sidewaysfigure}

\clearpage
\begin{sidewaysfigure}
\includegraphics[width=18cm]{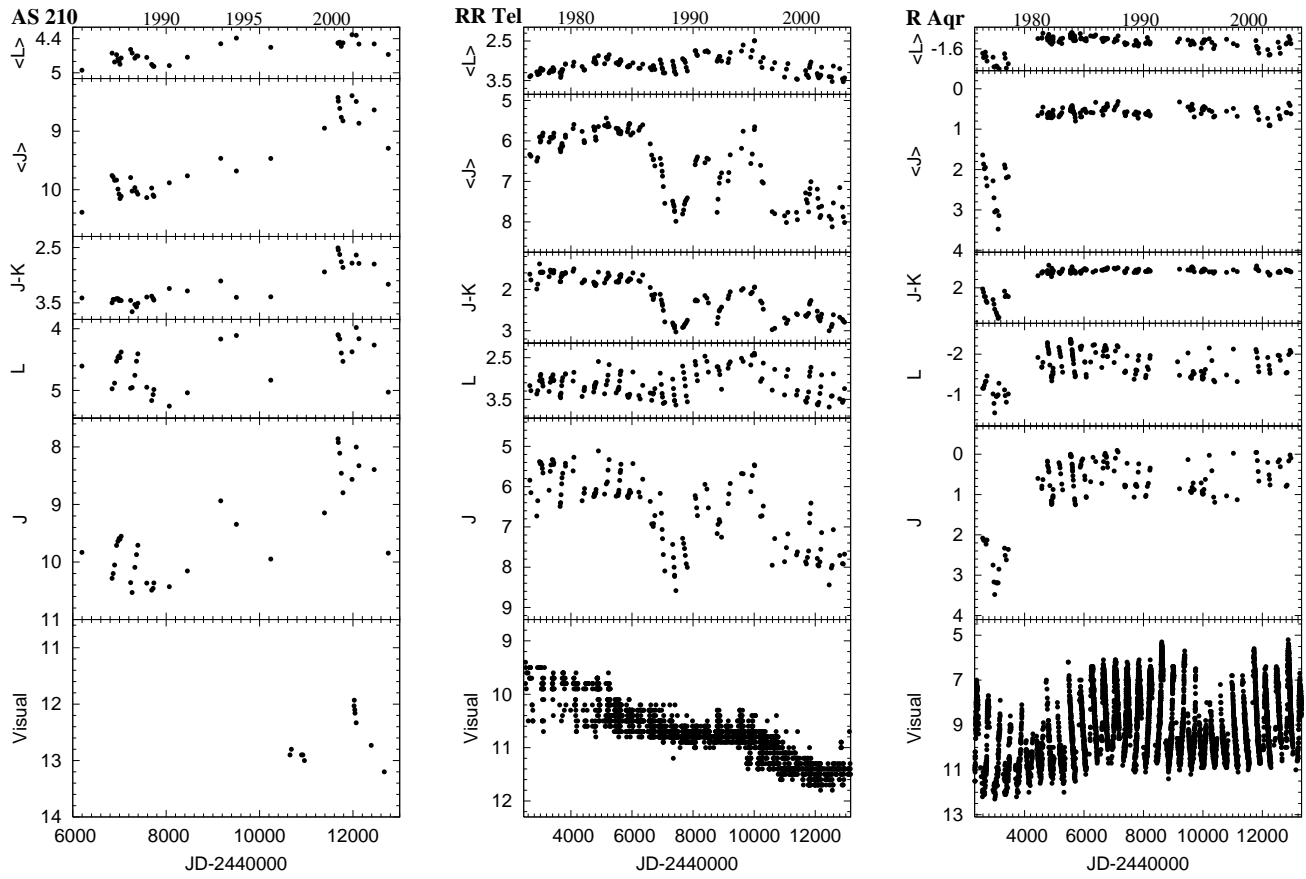}
\caption{Near-IR light curve of AS 210 (left panels), RR Tel (middle panels) and R Aqr (right panels).}
\end{sidewaysfigure}

\newpage
\begin{figure}
\begin{center}
\includegraphics[width=11.5cm]{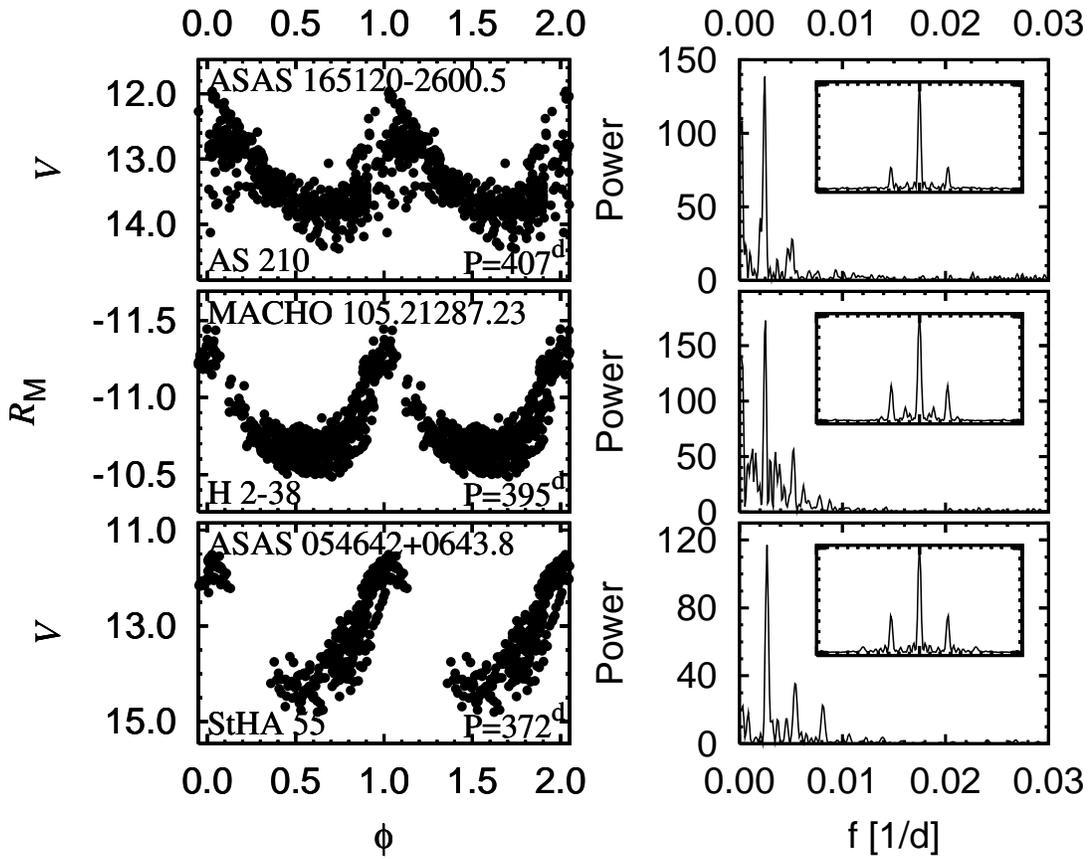}
\end{center}
\caption{Light curves of AS~210, H~2-38 and StHA~55 folded with pulsation periods (left panels) and related power spectra (right panels). Insights in top right corners of power spectra panels show power spectrum of windows.}
\end{figure}

\newpage
\begin{figure}
\begin{center}
\includegraphics[width=11.5cm]{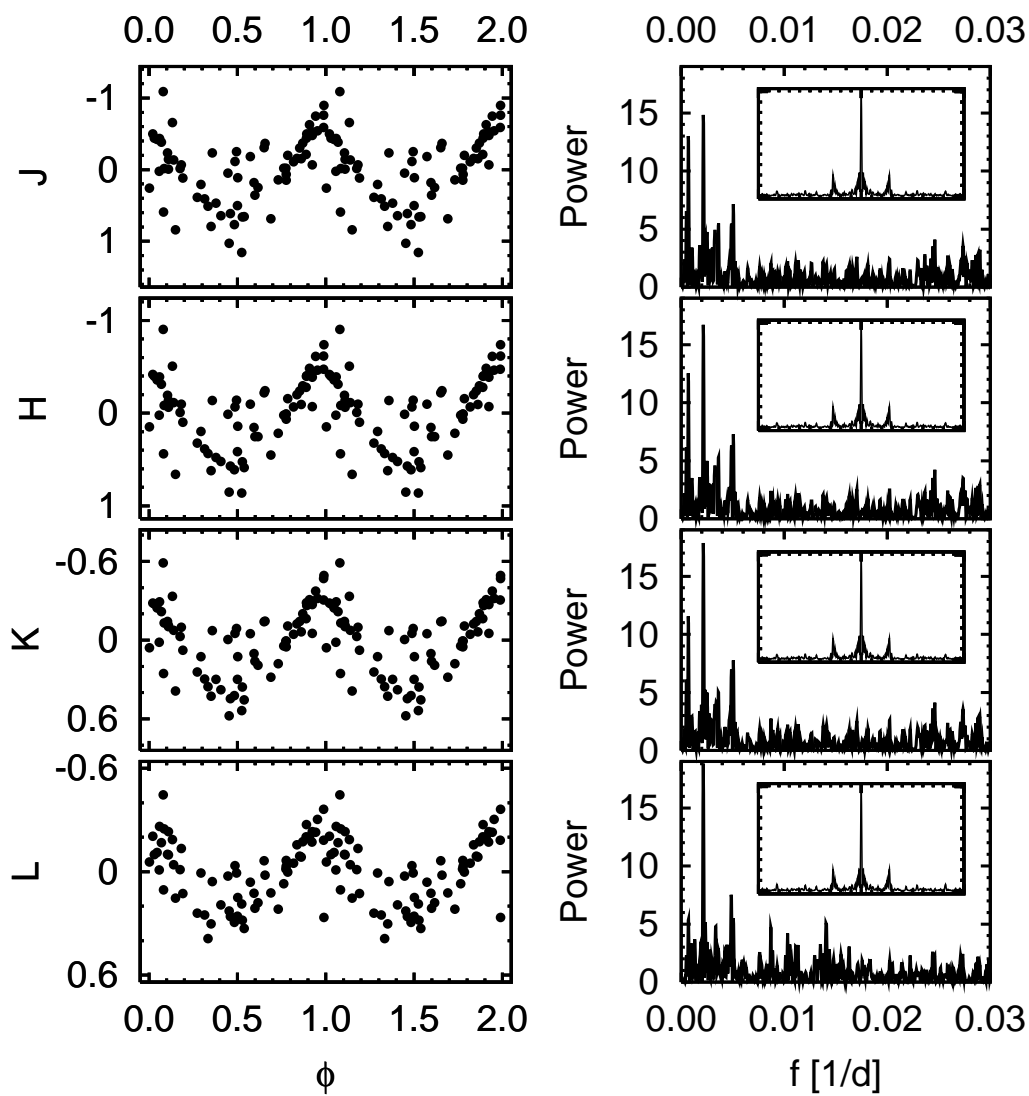}
\end{center}
\caption{Near-IR light curves of SS73~38 folded with pulsation periods (left panels) and related power spectra (right panels). Insights in top right corners of power spectra panels show power spectrum of windows.}
\end{figure}

\clearpage
\newpage
\begin{figure}
\includegraphics[width=12cm]{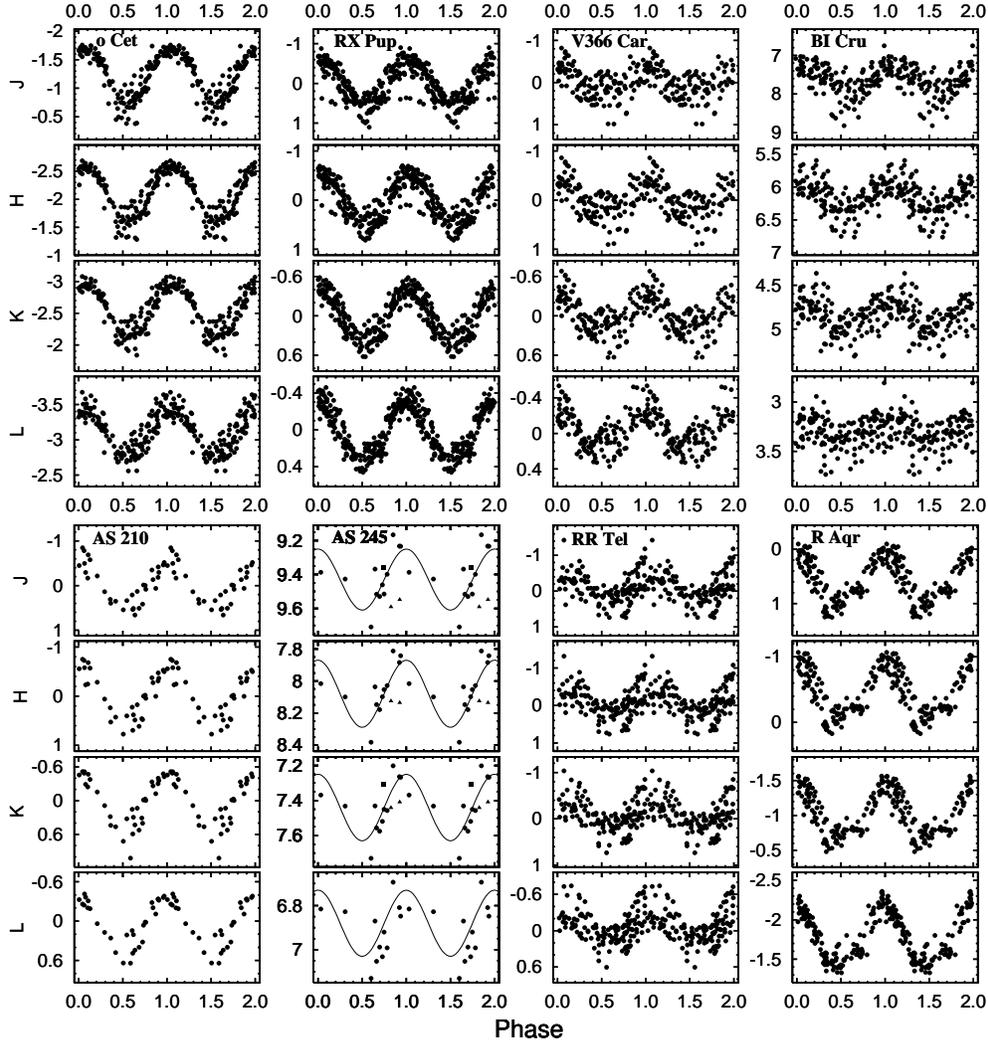}
\caption{Near-IR light curves of studied objects folded with pulsation periods. Ephemerides from Table 4 are used
for most objects with the exception of $o$~Cet and R~Aqr. In case of these object ephemerides are taken from Kholopov 1985. In panels of AS~245 dots represent SAAO data, triangle 2MASS, squares DENIS.}
\end{figure}

\newpage
\begin{figure}
\begin{center}
\includegraphics[width=11.5cm]{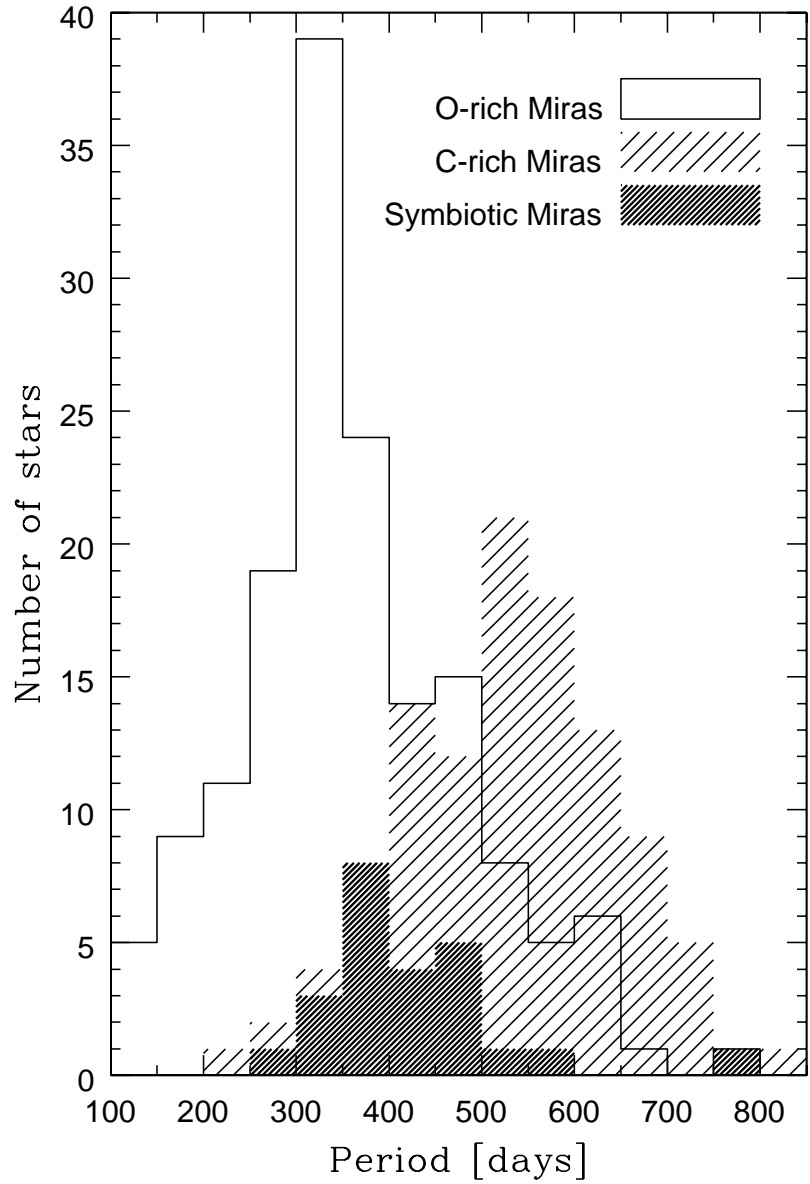}
\end{center}
\caption{The pulsation period distribution for symbiotic Miras together 
with those for normal field Miras (Olivier \etal
2001, Whitelock \etal 1994, 2000, 2006). }
\end{figure}

\clearpage

\newpage
\begin{figure}
\includegraphics[width=11.5cm]{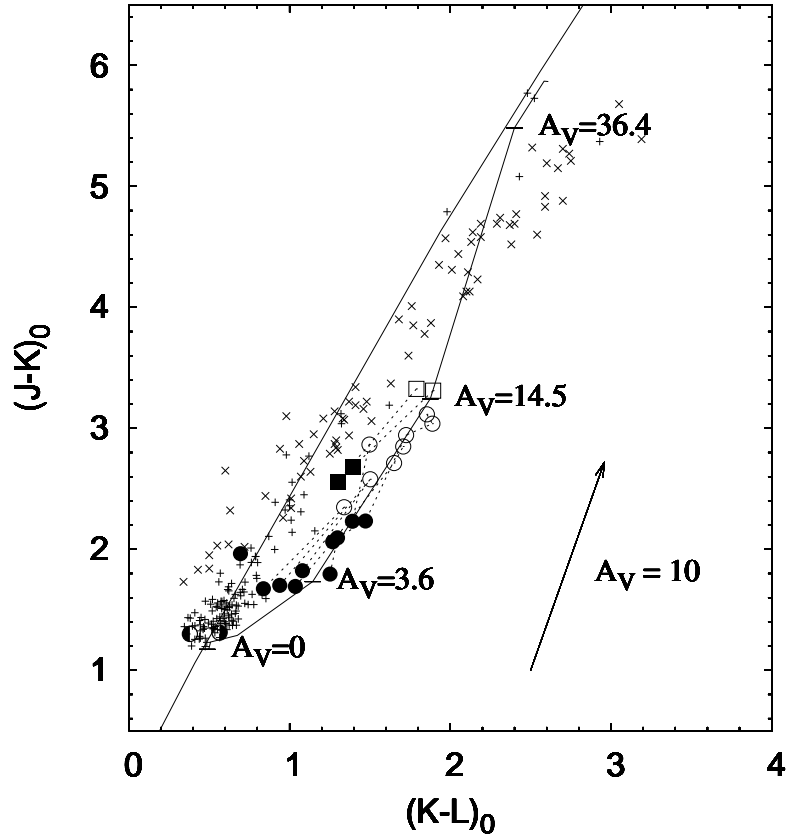}
\caption{The $(J-K)_0$ vs. $(K-L)_0$ colour-colour diagram. Colours are deredded using
extinction from galactic 3-D model. 
Open dots represent oxygen rich symbiotic Miras during dust obscuration, filled dots outside obscuration.
Open squares represent carbon rich symbiotic Miras during dust obscuration, whilst filled outside obscuration.
For comparison we also plot colours of oxygen rich Miras (+) and
carbon rich Miras ($\times$). For objects with uncertain nature different 
symbols were used: $o$~Cet  ($\RIGHTcircle$), AS~245 ($\LEFTcircle$).
Straight line shows black body colours. 
Simple model of star with shell dust is plot as well.
The temperature of star is 2750 K (black body), whilst dust shell temperature is 800 K. 
This model also includes line blanketing. }
\end{figure}

\newpage
\begin{figure}
\begin{center}
\includegraphics[width=11.5cm]{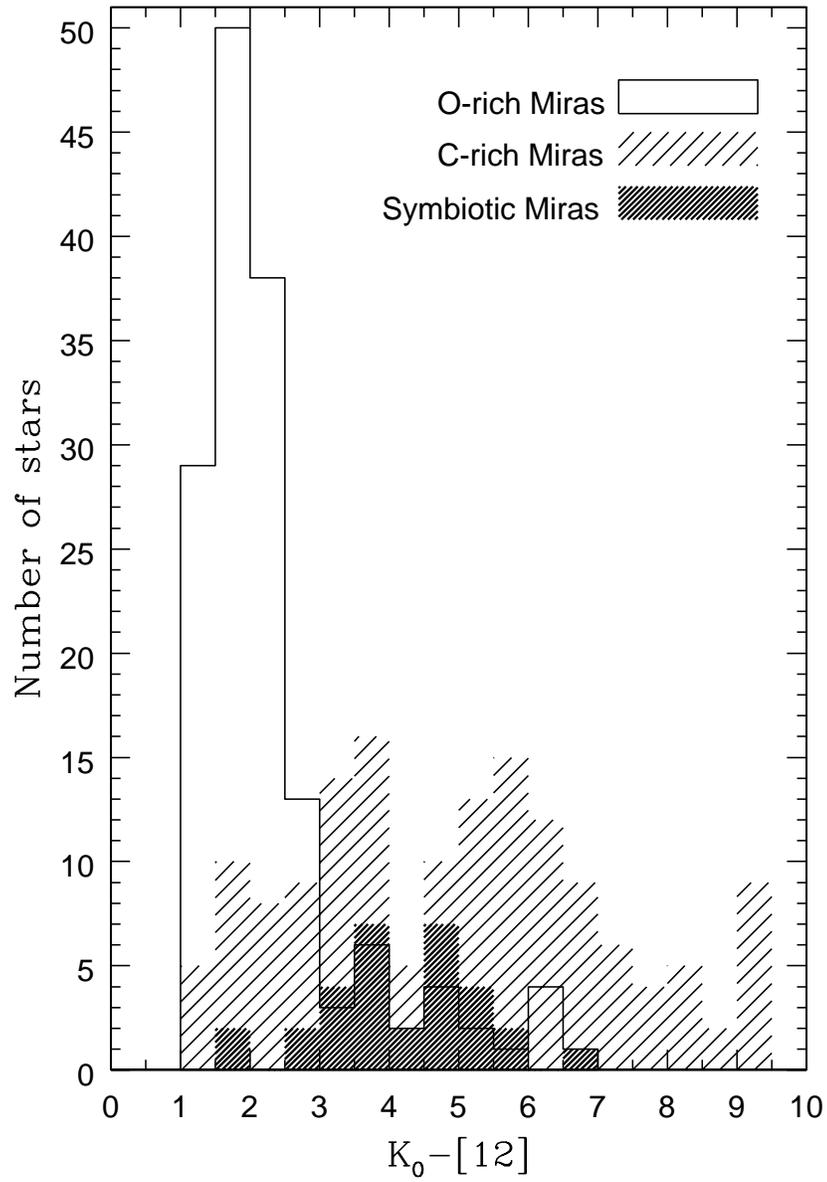}
\end{center}
\caption{$K_0-[12]$ distribution for symbiotic Miras together with those for
normal Miras.}
\end{figure}

\newpage
\begin{figure}
\begin{center}
\includegraphics[width=11.5cm]{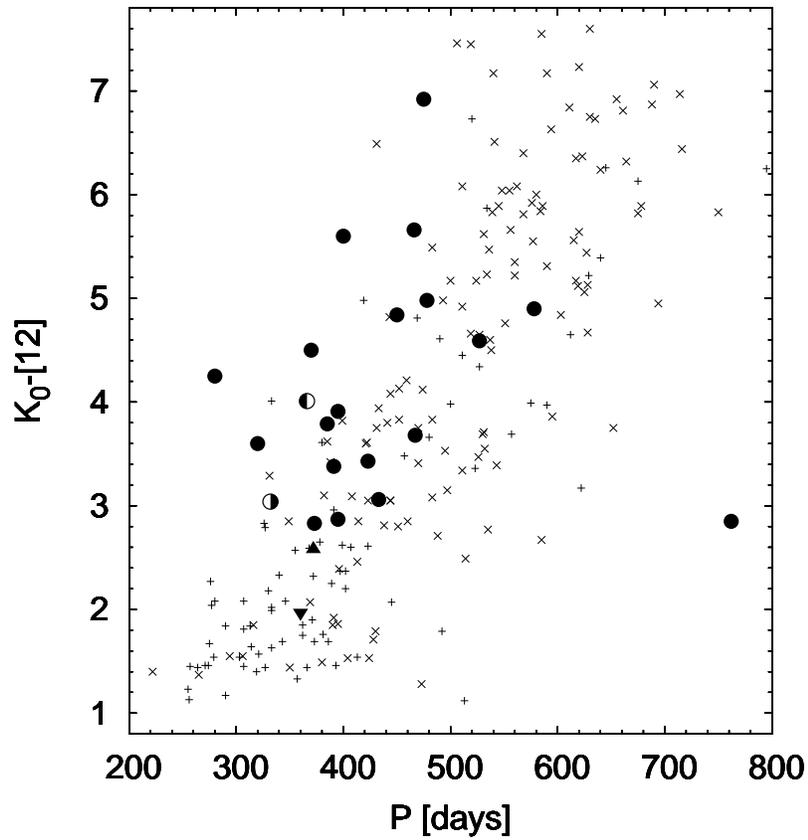}
\end{center}
\caption{Pulsation period vs. $K_0-[12]$ colour for symbiotic Miras ($\bullet$). 
For comparison we also plot colours of oxygen rich Miras (+) and
carbon rich Miras ($\times$). For objects with uncertain nature different 
symbols were used: $o$~Cet  ($\RIGHTcircle$), AS~245 ($\LEFTcircle$), 
StHA~55 ($\blacktriangle$), and V335~Vul ($\blacktriangledown$). The
descrepant point at P=763 days is V407 Cyg, which is thought to be hot
bottom burning giant and significantly more massive than the other symbiotic Miras
(see text for detail).}
\end{figure}

\newpage
\begin{figure}
\begin{center}
\includegraphics[width=11.5cm]{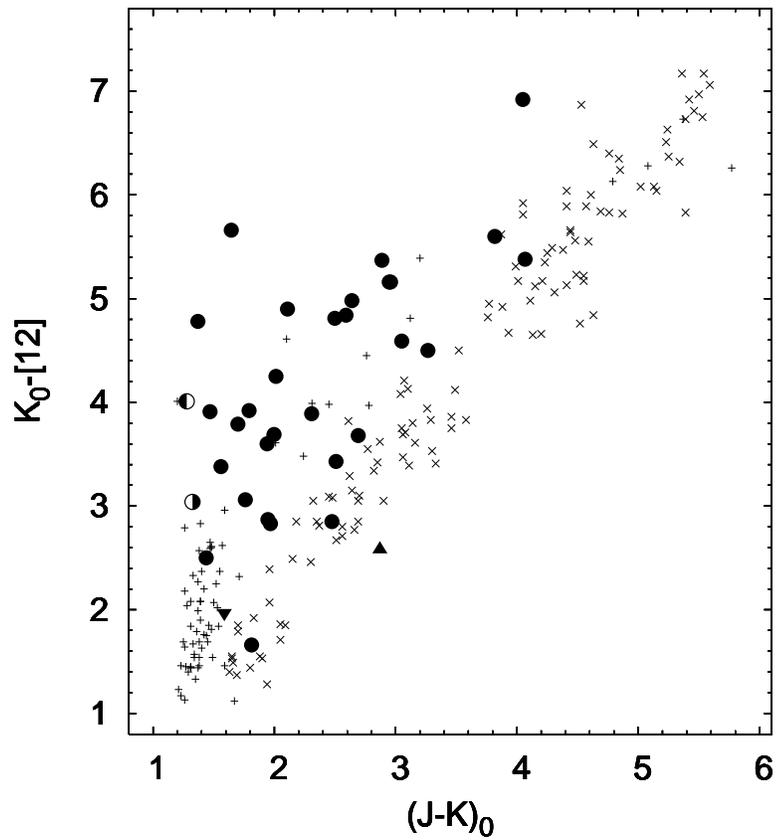}
\end{center}
\caption{$(J-K)_0$ vs. $K_0-[12]$ colour for symbiotic Miras ($\bullet$). 
For comparison we also plot colours of oxygen rich Miras (+) and
carbon rich Miras ($\times$). For objects with uncertain nature different 
symbols were used: $o$~Cet  ($\RIGHTcircle$), AS~245 ($\LEFTcircle$), 
StHA~55 ($\blacktriangle$), and V335~Vul ($\blacktriangledown$).}
\end{figure}

\end{document}